\documentclass[aps,twocolumn,epsfig,eqsecnum,showpacs]{revtex4}
\usepackage{amsthm}
\usepackage{amsmath}
\usepackage{amsfonts}
\usepackage{amssymb}

\newcommand\be{\begin{eqnarray}}
\newcommand\ee{\end{eqnarray}}
\newcommand\ba{\begin{array}}
\newcommand\ea{\end{array}}

\newtheorem{theorem}{Theorem}
\newtheorem{lemma}{Lemma}

\theoremstyle{definition}

\def\r{\rangle}
\def\l{\langle}
\def\T{{\rm Tr}}

\def\cH{{\cal H}}
\def\cS{{\cal S}}

\def\cR{{\cal R}}

\def\cI{{\cal I}}

\def\cT{{\cal T}}

\def\cE{{\cal E}}

\def\cM{{\cal M}}

\def\cA{{\cal A}}
\def\cB{{\cal B}}

\begin{document}
\title{Process positive operator valued measure: A mathematical framework for the description of process tomography experiments}
\author{M\'ario Ziman$^{1,2}$}
\affiliation{
$^{1}$Research Center for Quantum Information, Slovak Academy of Sciences,
D\'ubravsk\'a cesta 9, 845 11 Bratislava, Slovakia \\
$^{2}$ Faculty of Informatics, Masaryk University, Botanick\'a 68a,
602 00 Brno, Czech Republic
}
\begin{abstract}
In this paper we shall introduce the mathematical framework for
the description of measurements of quantum processes. Using this framework
the process estimation problems can be treated in the similar way as the state
estimation problems, only replacing the concept of positive operator valued 
measure (POVM) by the concept of
process POVM (PPOVM). In particular, we will show that any measurement 
of qudit channels 
can be described by a collection of effects (positive operators) defined 
on two-qudit system. However, the effects
forming a PPOVM are not normalized in the usual sense. We will demonstrate the 
usage of this formalism in discrimination problems by showing that
perfect channel discrimination is equivalent to a specific unambiguous
state discrimination. 
\end{abstract}

\maketitle


\section{Introduction}
Quantum and classical information processing are both based on quantum
properties of matter and light. Therefore, a big experimental
effort is paid in order to increase our abilities to control and manipulate 
individual quantum systems. Depending on the particular problem 
the goal of quantum experiments 
is to design a quantum device either able to prepare a quantum state, or
to perform a measurement, or to implement a quantum process. 

In quantum theory the states and the measurements are intimately 
related via the Born's duality relation predicting the quantum probabilities
\cite{perez,busch}. In particular, for a system prepared in a state 
$\varrho\in\cS(\cH)=\{\varrho\ge 0,\T\varrho=1\}$
the probability of measuring the outcome 
associated with an effect $F$ ($0\le F\le I$) is defined as 
$p(F|\varrho)=\T\varrho F$. Consequently, the measurement devices 
giving rise to outcomes $x_j$ are described 
by collections of quantum effects $F_j$ forming the so-called
positive operator valued measure (POVM), i.e. $\sum_j F_j=I$. 
Most of the problems (such as state estimation \cite{helstrom,holevo,rehacek}, 
state discrimination \cite{helstrom,ivanovic,chefles}, 
state comparison \cite{barnett}, etc.) related to the identification 
of quantum states can be mathematically formulated in the language 
of POVMs.

However, the preparation of states is not the only interesting 
experimental task. For example, the implementation of specific quantum 
processes is one of the main goal of the area of quantum information processing 
\cite{nielsen} aiming to run useful quantum algorithms.
The identification problems for states (preparators) can be 
naturally extended to processes. But, a fundamental concept
playing the role of POVM is missing. The main aim of this paper 
is to introduce a resembling mathematical 
framework for the description of all possible experiments 
measuring the quantum channels. This framework 
will simplify the investigation of the process identification 
problems.

\section{Measurements of channel parameters}
Consider an unknown quantum channel $\cE$ (i.e. a completely positive 
tracepreserving linear map 
\cite{davies}) acting on a $d$-dimensional quantum 
system (qudit). The most general {\it process/channel measurement} $\cM$
consists of the following three steps:
\begin{enumerate}
\item{}
Preparation of a (test) state $\varrho_j\in\cS(\cH_{\rm anc}\otimes\cH)$ 
of $D_j\times d$- dimensional system, thus, initially the testing system 
is composed of a qudit and an ancillary system of dimension $D_j$.
The ancillary system can be of different size for a different test 
state $\varrho_j$.
\item{}
Application of an unknown process $\cE$ on the qudit and some known 
channel $\cT_{j,\rm anc}$ on the ancillary quantum system.
\item{}
A measurement $M_j$ (given as a collection of 
positive operators, $F_{jk}\ge 0$, 
summing up to identity operator, $\sum_k F_{jk}=I$
for all $j$) of the output state 
$\varrho^\prime_j=(\cT_{j,\rm anc}\otimes\cE)[\varrho_j]$
results in an outcome $k$ with a probability
$p_{jk}(\cE)={\rm Tr}[F_{jk}\varrho_j^\prime]$.
\end{enumerate}

It follows that a general experiment measuring a process $\cE$
is associated with a collection of triples 
$\cM_{jk}=\l\varrho_j,\cT_{j,\rm anc},F_{jk}\r$
occurring with probabilities $p_{jk}$ defined above.
However, a channel $\cT_{j,\rm anc}$ can be considered as being 
a part of a preparation, or a measurement process, i.e. the 
triples $\l\varrho_j,\cT_{j,\rm anc},F_{jk} \r$,
$\l\cT_{j,\rm anc}\otimes\cI[\varrho_j],\cI_{\rm anc},F_{jk}\r$, and
$\l\varrho_j,\cI_{\rm anc},\cT^*_{j,\rm anc}\otimes\cI[F_{jk}] \r$,
(where $\cI$ is the identity quantum channel, and
$\cT_{j,\rm anc}^*$ is defined via the duality relation
$\T \{B^\dagger\cT_{\rm anc}[A]\}=\T\{(\cT^*_{\rm anc}[B])^\dagger A\}$
holding for all operators $A,B$)
define the same probabilities $p_{jk}(\cE)$.
Without the loss of generality 
we may assume that the ancilla system evolves trivially,
$\cT_{j,\rm anc}=\cI_{\rm anc}$ for all $j$, 
hence, the triples can be replaced by couples 
$\cM_{jk}=\l\varrho_j,F_{jk}\r$ occurring with probabilities 
$p_{jk}(\cE)=\T \{(\cI_{\rm anc}\otimes\cE)[\varrho_j] F_{jk}\}$.

The following lemma is a version of the so-called Choi-Jamiolkowski
isomorphism \cite{choi,jamiolkowski}
relating qudit linear maps with linear operators on $d\times d$ system.

\begin{lemma}
For arbitrary state of $D\times d$ system 
($\varrho\in\cS(\cH_D\otimes\cH_d)$)
there exists a completely positive linear map
$\cR_\varrho:\cB(\cH_d)\to\cB(\cH_D)$ such that
$(\cR_\varrho\otimes\cI)[\Psi_+]=\varrho$, where 
$\Psi_+=|\Psi_+\r\l\Psi_+|$ and $|\Psi_+\r=\sum_{j=1}^d |j\r\otimes |j\r$
is an unnormalized maximally entangled
quantum state on $d\times d$ system.
\end{lemma}

\begin{proof}
Consider a pure state $\varrho=|\phi\r\l\phi|=\Phi$ of 
a $d\times D$-dimensional system and 
$|\phi\r=\sum_{j=1}^d\sum_{\alpha=1}^D \Phi_{\alpha j}|\alpha\r\otimes|j\r$.
Define an operator $A_\Phi:\cH_d\to\cH_D$ acting as 
follows $A_\Phi\otimes I_d|\Psi_+\r=|\Phi\r$, i.e.
$A_\phi|j\rangle=\sum_{\alpha=1}^D \Phi_{\alpha j}|\alpha\r$. 
We can write 
$\Phi=(\cR_\Phi\otimes\cI)[\Psi_+]=
(A_\Phi\otimes I_d)\Psi_+(A^\dagger_\Phi\otimes I_d)$,
where $\cR_\Phi$ is a unique linear completely positive map,
because the expression
of $|\Phi\r$ in the basis $\{|\alpha\r\otimes|j\r\}$ is unique. 
Generalization to an arbitrary mixed state is straightforward.
For $\varrho=\sum_j \lambda_j |\phi_j\r\l\phi_j|$ we can define a map
$\cR_\varrho=\cR_{\sum_j \lambda_j \Phi_j}=\sum_j \lambda_j \cR_{\Phi_j}$
that maps the maximally entangled state $\Psi_+$ into 
$\varrho=(\cR_\varrho\otimes\cI)[\Psi_+]=\sum_j \lambda_j 
(\cR_{\Phi_j}\otimes\cI)[\Psi_+]=\sum_j\lambda_j\Phi_j$. Different
convex decompositions of $\varrho$ into pure states define different
Kraus decompositions of the same completely positive map $\cR_\varrho$,
hence, this map is unique. It is straightforward to see that similar
result holds for a general positive operator $F$, that is, the
transformation $\cR_F$ defined as $\cR_F\otimes\cI[\Psi_+]=F$ 
is completely positive, too.
\end{proof}

Using this lemma and the definition of the dual map $\cR^{*}_{\varrho}$  
the probability for a couple $\l\varrho,F\r$ of the test state
$\varrho$ and a measurement resulting in an outcome associated with
an effect $F$ can be expressed as follows
\be
\nonumber
p(\cE)&=&\T\{(\cI_{\rm anc}\otimes\cE)[\varrho]F\}\\ \nonumber
&=&\T\{(\cR_{\varrho}\otimes\cE)[\Psi_+]F\}\\ \nonumber
&=& \T\{(\cI_{\rm anc}\otimes\cE)[\Psi_+](\cR_{\varrho}^*\otimes\cI)[F]]\}\\
\nonumber &=&\T\{(\cI_{\rm anc}\otimes\cE)[\Psi_+]\cM\}\, .
\ee
Based on this calculation we see that
an operator $\cM=(\cR_{\varrho}^*\otimes\cI)[F]$ 
completely describes the considered process measurement outcome associated with
$\l\varrho,F\r$. Since both the operations $\cR_{\varrho},\cR_{\varrho}^*$ 
are completely positive, but not necessarily trace-preserving,
it follows that an operator 
\be
\cM=(\cR_\varrho^*\otimes \cI)[F]
\ee 
is positive and $\cM\le I_{d\times d}$, that is, $\cM$ is an effect defined 
on a $d\times d$-dimensional system that we shall call a
{\it process/channel effect}.
The most general {\it process/channel measurement} is defined
as a collection of process effects 
$\cM_{jk}=p_j(\cR_{\varrho_j}^*\otimes\cI)[F_{jk}]$ associated with
couples $\l p_j\varrho_j, F_{jk}\r$ with $\sum_k F_{jk}=I_{D\times d}$ 
for all $j$ labeling potentially different test states $\varrho_j$ 
chosen with a prior distribution $p_j$. 

Let us assume that the process is probed only by a single test state 
$\varrho$, i.e. $\cM_k\leftrightarrow\l\varrho,F_k\r$. In such case
$\sum_{k} \cM_{k}=(\cR_{\varrho}^*\otimes\cI)[\sum_{k} F_{k}]
=(\cR_{\varrho}^*\otimes\cI)[I_{D\times d}]$. Since 
$\cR_{\varrho}[X]=\sum_j \lambda_j A_{\Phi_j}XA_{\Phi_j}^\dagger$
the action of the dual map can be expressed as 
$\cR^*_{\varrho}[X]=\sum_j \lambda_j A^\dagger_{\Phi_j}XA_{\Phi_j}$.
Consequently, we get that the following normalization condition holds
\be
\nonumber
\sum_{k} \cM_{k}=\sum_j \lambda_j A^\dagger_{\Phi_j}A_{\Phi_j}\otimes I_d
=(\T_{\rm anc}\varrho)^T\otimes I_d\, .
\ee
Thus, the process effects $\cM_{k}$ form a positive operator valued measure
not necessarily normalized in the usual sense, because 
$\sum_k \cM_k\le I_{d\times d}$.

For a general process measurement (described by process effects
$\cM_{jk}=p_j(\cR_{\varrho_j}^*\otimes\cI)[F_{jk}]$) 
it follows that
$\sum_{jk}\cM_{jk}= \sum_j(p_j\T_{\rm anc}\varrho_j)^T \otimes I_d
=(\T_{\rm anc}\overline{\varrho})^T\otimes I_d$,
where the operator $\overline{\varrho}=\sum_j p_j\varrho_j$ 
is the average test state. Even if the test states are using 
different ancillas, it is always possible
to consider them as joint states of the qudit and the largest 
of the ancilla systems. It follows that
the process measurement consisting of process effects
$\cM_{jk}\leftrightarrow\l p_j\varrho_j,F_{jk}\r$ 
can be understood as a process measurement
composed of $\cM_{jk}\leftrightarrow\l\Xi,|j\r\l j|\otimes F_{jk}\r$ 
with a single test state $\Xi=\sum_{j} p_j|j\r\l j|\otimes\varrho_j$. 

We have shown that each qudit channel measurement can be associated with
a {\it process positive operator valued measure} (PPOVM), i.e.
by a collection of effects $\cM_\alpha$ of $d\times d$-dimensional system
summing up to $\varrho^T\otimes I_d$, where $\varrho$ is a qudit quantum
state. In the following we will prove that the converse of this statement
also holds. 

\begin{theorem}
Each PPOVM can be implemented as a process measurement.
\end{theorem}
\begin{proof}
Consider a PPOVM $\{\cM_\alpha\}$ with
$\sum_{\alpha}\cM_\alpha=\varrho^T\otimes I_d$. Our aim is to show that
this PPOVM really corresponds to a process measurement. As it was argued
before we can restrict ourselves to a process measurement using 
only a single test state $\Xi$ such that $\T_{\rm anc}\Xi=\varrho$.
Moreover, assuming that the test state is a pure state, 
the question is whether
$\cM_\alpha=(A_\Xi^\dagger\otimes I_d)F_\alpha(A_\Xi\otimes I_d)$ implies
that 
\be
\nonumber
F_\alpha=([A_\Xi^\dagger]^{-1}\otimes I_d)\cM_\alpha(A_\Xi^{-1}\otimes I_d)\, ,
\ee
hence, whether the operators $A_\Xi,A_\Xi^\dagger$ are invertible.
Let $r={\rm rank}\varrho\le d$ and assume that $\Xi$
is a pure state of a qudit and an ancilla of the dimension $r$, hence,
$\cR_\Xi^*[X]=A_\Xi^\dagger X A_\Xi$ and
$A_\Xi^\dagger A_\Xi=(\T_{\rm anc}\Xi)^T=\varrho^T$.
The support of each operator $\cM_\alpha$ is a subset of the support
of $\varrho^T\otimes I$, i.e. both are defined on $(r\times d)$-dimensional
system. Since ${\rm rank}(A_\Xi^\dagger A_\Xi)={\rm rank}(A_\Xi A_\Xi^\dagger)=
{\rm rank}A_\Xi={\rm rank}A_\Xi^\dagger$ it follows that
operators $A_\Xi,A_\Xi^\dagger,\varrho^T,\varrho$ have the same rank
(equal to $r$). Because the operators $A_\Xi,A_{\Xi}^\dagger$  
act on $r$-dimensional ancilla system (they have full rank) 
it follows they are invertible.
Consequently, the above equation defines
positive operators $F_\alpha$ forming a POVM, because
$\sum_\alpha F_\alpha=([A_\Xi^\dagger]^{-1}\varrho^T A_\Xi^{-1})\otimes I_d
=I_r\otimes I_d$. 
\end{proof}

To summarize, we have shown that arbitrary collection of process
effects $\cM_\alpha$ forming PPOVM can be implemented
by using a pure state $|\Xi\r\in\cH_r\otimes\cH_d$ such that
$\T_{\rm anc}\Xi=\varrho$ and performing a POVM given by
positive operators 
$F_\alpha=([A_\Xi^\dagger]^{-1}\otimes I_d)\cM_\alpha(A_\Xi^{-1}\otimes I_d)$
with $A_\Xi=\sqrt{\varrho^T}$.  This result allows us to abstract 
particular experimental realizations of process measurements 
and employ the framework of PPOVM directly. In this framework
the qudit quantum channels are represented by positive two-qudit operators 
$\omega_\cE=\cI\otimes\cE[\Psi_+]$ satisfying $\T\omega_\cE=d$ and 
$\T_2\omega_\cE=I$. Let us denote the set of all processes 
(process state space) by $\cS_{\rm proc}=
\{\omega\in\cB_+(\cH\otimes\cH), \T_2\omega=I, \T\omega=d\}$. 
This set is convex 
and compact subset of the set of positive operators of trace $d$ denoted as
$\cB_{+d}(\cH\otimes\cH)$, which is isomorphic
to $\cB_{+1}(\cH\otimes\cH)=\cS(\cH\otimes\cH)$ 
(set of density matrices).

Let us illustrate the process POVM in two usually
analyzed process tomography experiments:
\begin{enumerate}
\item{\it Maximally entangled probe.}
Consider that an unknown qudit channel is probed 
by a (normalized) maximally entangled state
$|\psi_+\r=\frac{1}{\sqrt{d}}\sum_j |j\r\otimes|j\r$. In this case
the mapping $\cR_{\psi_+}=\frac{1}{d}\cI$, i.e. 
$|\psi_+\r\l\psi_+|=\frac{1}{d}\Psi_+$. That is, 
$\cM=(\cR_{\psi_+}^*\otimes\cI)[F]=\frac{1}{d}F$,
where $F$ is a two-qudit effect. Considering a POVM
consisting of effects $F_1,\dots,F_n$ the corresponding PPOVM
is composed of positive operators $\cM_j=\frac{1}{d}F_j$.

\item{\it Ancilla-free test states.}
In this case the qudit test state $\varrho$ can be 
understood as being a factorized state of an ancilla and a qudit, i.e.
$\Omega=\xi\otimes\varrho$. The POVM effects have the form 
$I_{\rm anc}\otimes F_j$ and the corresponding
process effects are $\cM_j=(\cR_\Omega^*\otimes\cI)
[I_{\rm anc}\otimes F_j]=\varrho^T\otimes F_j$. It follows that
if we want to perform an equivalent (defining the same PPOVM) 
process measurement with the maximally entangled probe, the 
the state measurement consists of positive operators 
$X_j=d\varrho^T\otimes F_j$. However, let us note that these 
operators are not effects.
\end{enumerate}

\section{Informationally complete process tomography}
A process measurement $\{\cM_\alpha\}$ is called informationally
complete if for each quantum process $\cE$ the probability distribution 
$p_\alpha(\cE)=\T[\omega_\cE\cM_\alpha]$ is different. Thus the process
can be uniquely identified from the observed probability distribution.
This happens if and only if a linear span of operators $\cM_\alpha$ 
contains the whole set of process states $\cS_{\rm proc}$.

Consider a qubit channel probed by a maximally entangled state
$|\psi_+\r=\frac{1}{\sqrt{2}}(|00\r+|11\r)$. Performing the measurements 
of sharp observables $\sigma_\mu\otimes\sigma_\nu$
(each with the probability 1/9), where $\mu,\nu=x,y,z$.
That is, the POVM is composed of effects 
$F_{a,b}=\frac{1}{9} |a\r\l a|\otimes |b\r\l b|=2\cM_{a,b}$,
where $a,b=\pm x,\pm y,\pm z$.
The states $|\pm x\r,|\pm y\r,|\pm z\r$ are the eigenvectors 
of $\sigma_x,\sigma_y,\sigma_z$ associated with eigenvalues $\pm 1$, 
respectively. The set of operators $\cM_{a,b}$ is overcomplete
and its span contains the whole set of process states, i.e.
it is an informationally complete PPOVM.  Calculating the sum we find that
$\sum_{a,b} \cM_{a,b}=\frac{1}{2}I_2\otimes I_2$.

Alternatively, one of the simplest experimental implementations of an
informationally complete process measurement
consists of the preparation of six test states $|\pm x\r,|\pm y\r,|\pm z\r$
distributed with the same probability $1/6$.
The measurement of the output states is the complete qubit state
tomography measuring all three Pauli operators $\sigma_x,\sigma_y,\sigma_z$,
hence, it consists of effects 
$F_{\pm a}=\frac{1}{3}|\pm a\r\l\pm a|$ ($a=x,y,z$). Consequently, the whole
setup is described by PPOVM composed of operators
$\cM_{\nu,\mu}=\frac{1}{18}|\nu\r\l\nu|^T\otimes|\mu\r\l\mu|$
with $\sum_{\nu,\mu}\cM_{\nu,\mu}=\frac{1}{2}I_2\otimes I_2$,
where $\nu,\mu=\pm x,\pm y,\pm z$. Let us note that PPOVMs
$\{\cM_{\mu,\nu}\}$ and $\{\cM_{a,b}\}$ (described in the 
previous paragraph) coincide, because 
$(|\pm x\r\l\pm x|)^T=|\pm x\r\l\pm x|$,
$(|\pm y\r\l\pm y|)^T=|\mp y\r\l\mp y|$,
$(|\pm z\r\l\pm z|)^T=|\pm z\r\l\pm z|$, where the transposition
is performed with respect to basis $|\pm z\r$.

\section{Perfect discrimination}
Two processes $\cE_1,\cE_2$ are perfectly distinguishable
if there exists an experimental setup such that in its single
run the outcomes uniquely identify the process. It corresponds
to an existence of a two-outcome PPOVM, $\cM_1+\cM_2=\varrho^T\otimes I$,
such that $p_1(\cE_1)p_1(\cE_2)=p_2(\cE_1)p_2(\cE_2)=0$. That is,
the process effect $\cM_1$ is associated with the conclusion
that the process is $\cE_1$, and the process effect 
$\cM_2$ corresponds to the conclusion 
$\cE_2$. The conditions
$\T \cM_1\omega_{\cE_2}=0$ and $\T\cM_2\omega_{\cE_1}=0$ imply
that ${\rm supp}[\cM_{1}]\perp{\rm supp}[\omega_{\cE_2}]$ and 
${\rm supp}[\cM_{2}]\perp{\rm supp}[\omega_{\cE_1}]$, where $\omega_{\cE_1}$
and $\omega_{\cE_2}$ are the corresponding process states. 
Without any doubts the process and state discrimination tasks
are closely related and it seems they are almost the same in the sense 
that process discrimination problems are reducible to state discrimination 
problems. It is so indeed, but there is still one important difference: 
PPOVMs are not normalized to identity. As the consequence of this fact 
we cannot make a conclusion that orthogonality of supports of
$\omega_{\cE_1}$ and $\omega_{\cE_2}$ is the necessary condition for 
perfect discrimination of processes $\cE_1$ and $\cE_2$. 
In fact, there are process states with non-orthogonal 
supports that can be perfectly discriminated by means of PPOVM. 

In particular, consider one of the channels being 
the identity map ($\cE_1=\cI$) and second one
transforming the whole state space into
a fixed pure state $|0\r$ ($\cE_2=\cA_0$). The corresponding
operators $\omega_\cI=\Psi_+$, $\omega_0=I\otimes|0\r\l 0|$, have non-orthogonal
supports, i.e. if considered as states they are not perfectly
distinguishable. However, there exists a very simple experimental
procedure of channel discrimination using the test state $|1\r$.
Probing the identity the output state is $|1\r$, whereas probing 
the contraction $\cA_0$ the output state is $|0\r$, i.e. which is
orthogonal to $|1\r$. A simple measurement (described by POVM elements 
$|0\r\l 0|, I-|0\r\l 0|$) tells us whether the channel was $\cI$, or $\cA_0$.
The corresponding PPOVM consists of process effects
$\cM_\cI=|1\r\l 1|\otimes (I-|0\r\l 0|)$ and 
$\cM_0=|1\r\l 1|\otimes|0\r\l 0|$,
$\cM_\cI+\cM_0=|1\r\l 1|\otimes I$. It is straightforward to verify that
$\T\cM_\cI \omega_\cI=\T\cM_0\omega_0=1$.

The characterization of all channels that can be perfectly discriminated
is beyond the scope of this paper. Instead we will provide qualitative
arguments why the orthogonality of supports is only sufficient, but not
necessary for perfect distinguishability of processes. In a sense any 
PPOVM can be understood as a normalized POVM (of two qudits 
states of trace $d$) if an effect $\cM_{\rm extra}=
I_d\otimes I_d-\sum_\alpha\cM_\alpha=(I_d-\varrho^T)\otimes I_d$ is added.
Because of the different normalization of states the
probabilities given by the trace relation are normalized to $d$
and we will use the term ``rate'' instead of ``probability''.
In particular, the rate to get the extra outcome equals
${\rm Tr}\omega\cM_{\rm extra}=d-1$ for all process states $\omega$.
This ``extra'' outcome is a fake outcome that is not really measured 
in the process measurement, but formally it describes the outcome of 
a process state measurement for which 
no conclusion is made. That is, the perfect discrimination of processes
by means of PPOVM can be understood as a special case of unambiguous 
discrimination of states via POVM. The inconclusive
result is associated with the "extra'' outcome added to PPOVM. 
And in such case the orthogonality of supports is not required. 
In particular, the process states $\omega_0,\omega_\cI$ (defined above) 
can be unambiguously discriminated.

\subsection{Perfect discrimination of unitary channels}
The discrimination of unitary processes was already investigated
in several papers \cite{acin,dariano,duan_2006} and the
solution is, in principle, known.
In the framework of PPOVMs the unitary channels are associated 
with maximally entangled process states $|\omega_U\r=U\otimes I|\Psi_+\r$.
According to previous section a pair of unitary channels $U,V$ can be perfectly
discriminated if and only if we can design an unambiguous state discrimination
machine of process states $\omega_U=|\omega_U\r\l\omega_U|$ 
and $\omega_V=|\omega_V\r\l\omega_V|$. For the inconclusive result
$\cM_{\rm extra}=(I_d-\varrho^T)\otimes I_d$ it is guaranteed that
\be
\l\omega_U|\cM_{\rm extra}|\omega_U\r=\l\omega_V|\cM_{\rm extra}|\omega_V\r
=d-1\, .
\ee
For the total failure rate we get 
$r_{\rm failure}={\rm Tr}[\cM_{\rm extra}(p_U\omega_U+p_V\omega_V)]=(d-1)$.
As it was shown in \cite{chefles} in the optimal case 
the failure probability of the discrimination between 
two pure states is given by the absolute value
of their scalar product, i.e.
$P_{\rm failure}(|\psi\r,|\varphi\r)=|\l\psi|\varphi\r|$. If applied for
process states $|\omega_U\r,|\omega_V\r$ (normalized as 
$\T\omega_U=\T\omega_V=d$) 
we see that the failure rate is $r_{\rm failure}=|\l\omega_U|\omega_V\r|=
|\T U^\dagger V|$. Therefore, whenever $|\T U^\dagger V|>d-1$ it follows
that the unitary channels $U$ and $V$ cannot be perfectly discriminated.
In other words, the inequality $|\T U^\dagger V|\le d-1$ is a 
necessary condition for perfect discrimination of unitary channels.

The problem of perfect discrimination of two unitaries
is equivalent to the discrimination of a single unitary channel and
the identity channel, $U,I$. Process effects $\cM_U,\cM_I$
are related to POVM consisting of two projectors $E_I,E_U$
via the relation $\cM=(\cR_\Omega^*\otimes\cI)[E]$, where the test state
$\Omega$ can be chosen to be pure. Identity does not affect
this state and therefore $E_I=|\Omega\r\l\Omega|=\Omega$ 
is the effect identifying the identity operator. Consequently, 
$E_U=I-\Omega$ is associated with the unitary channel $U$. 
The no-error condition $\l\omega_U|\cM_I|\omega_U\r=0$ 
and the definition of $\cM_I
=(A_\Omega^\dagger\otimes I)E_I(A_\Omega\otimes I)$ 
result in identity
\be
\nonumber
0=|\l\Psi_+|A_\Omega^\dagger A_\Omega\otimes U|\Psi_+\r|^2
=|\l\Omega|I\otimes U|\Omega\r|^2\, ,
\ee
hence, the existence of perfect discrimination is
guaranteed if and only if there exists a pure state $\Omega$
such that $\l\Omega|I\otimes U|\Omega\r$ vanishes. As it was argued
in \cite{acin,dariano,duan_2006} this is possible if and only
if zero belongs to a convex hull of eigenvalues of $U$ distributed
on a unit circle in the complex plane. For a general pair of unitaries
$U,V$ the problem is reduced to the analysis of eigenvalues of $UV^\dagger$.

In the qubit case each unitary has two eigenvalues, thus the perfect
discrimination of a pair $I,U$ requires that 
$U=e^{i\eta}|\varphi\r\l\varphi|+e^{i(\eta+\pi)}
|\varphi_\perp\r\l\varphi_\perp|$, i.e. ${\rm Tr} U=0$. Consequently,
qubit unitary channels $U,V$ can be perfectly discriminated 
if and only if they are orthogonal. However, such statement no longer
holds for qudits and as it was shown
in \cite{acin} for an arbitrary pair of (qudit) unitary processes $U,V$ 
there exists a finite $n$ such that $U^{\otimes n}$ and $V^{\otimes n}$ 
can be perfectly discriminated, i.e. 
the distinguishability is not equivalent to the orthogonality.
Let us note that this qubit example also shows that the necessary
condition $|\T U^\dagger V|\le d-1$ is not sufficient.

\section{Conclusions}
The goal of this paper has been to introduce a mathematical framework for
the description of measurements on quantum processes. This idea led us
to the definition of the so-called process POVM (PPOVM) defined as 
a collection of effects $\cM_1,\dots, \cM_n$ ($0\le \cM_j\le I_{d}\otimes I_d$),
such that $\sum_j \cM_j = \varrho^T\otimes I_d$, where $\varrho$ is an arbitrary
single qudit state and $^T$ denotes its transposition. In this framework the
channels are associated with positive operators of $d\times d$ system
with trace equal to $d$. An arbitrary process 
measurement can be described by PPOVMs and we have shown that also 
each PPOVM can be implemented experimentally although
the experimental realization is not unique. This ambiguity is one 
of important open problems left for further investigation.

The framework of PPOVMs provides us with a powerful tool 
for different process estimation problems 
\cite{acin,dariano,duan_2006,wang,dariano_sacchi_2005,li_qiu}, 
mostly in answering the optimality
questions. Moreover, the concepts originally developed for POVMs
can be directly translated and applied for PPOVMs as it was demonstrated
in the case of informational completeness of PPOVMs. Using the
PPOVM framework we have argued that perfect discrimination problems 
for quantum channels are equivalent to very specific unambiguous 
discrimination problems of quantum states. As it is discussed in 
\cite{dariano2008} the PPOVM framework is a special instance 
of a more general framework describing operations on quantum 
channels and observables.

{\it Acknowledgment.} This work was supported by the EU integrated
project QAP 2004- IST- FETPI-15848 and by Slovak grant agency
APVV RPEU-0014-06.


\end{document}